# Big Data links from Climate to Commodity Production Forecasts & Risk Management

Paulina Concha Larrauri[1], Upmanu Lall[1,2]

[1]Columbia Water Center, Columbia University, New York. [2] Earth & Environmental Eng., Columbia University, New York

Corresponding author: Paulina Concha Larrauri (pc2521@columbia.edu)

## Abstract

Frozen concentrated orange juice (FCOJ) is a commodity traded in the International Commodity Exchange. The FCOJ future price volatility is high because the world's orange production is concentrated in a few places, which results in extreme sensitivity to weather and disease. Most of the oranges produced in the United States are from Florida. The United States Department of Agriculture (USDA) issues orange production forecasts on the second week of each month from October to July. The October forecast in particular seems to affect FCOJ price volatility. We assess how a prediction of the directionality and magnitude of the error of the USDA October forecast could affect the decision making process of multiple FCOJ market participants, and if the "production uncertainty" of the forecast could be reduced by incorporating other climate variables. The models developed open up the opportunity to assess the application of the resulting probabilistic forecasts of the USDA production forecast error on the trading decisions of the different FCOJ stakeholders, and to perhaps consider the inclusion of climate predictors in the USDA forecast.

## 1. Introduction

The state of Florida is the largest producer of oranges in the United States and the second largest producer of orange juice in the world only next to Brazil (Hodges and Spreen, 2012). Around 90% of the oranges produced in Florida are processed into frozen concentrated orange juice (FCOJ)- traded in the International Commodity Exchange (ICE)[1]. The high concentration of most of the world's orange production in a few places results in an extreme sensitivity to weather and disease, as one extreme event such as a freeze or a hurricane can wipe out an entire crop (Miller et al., 1988). FCOJ future prices should consequently be affected mainly by such production risks, as demand fluctuations are relatively small (Chou et al, 2015). Surprisingly, a study by Roll (1984) showed that weather and weather forecasts explained only a small portion of the variability in FCOJ returns (calculated daily as price on day *t+1* minus price on day *t* divided by the price on day *t*). Boudoukh et al. (2007) revised the work of Roll (1984) and found that FCOJ prices do react to temperature in a significant way, and that approximately 50% of the variation in return can be explained by freezing temperatures. Of course other information about supply, such as news about weather, orange quality, Brazil's production, demand shifts, and production forecasts can also have important effects on prices. Chou et al., (2015) found that an investor-based measure such as the Consensus Bullish Sentiment Index (CBSI), which is specific to FCOJ, and temperature can explain around 20% of the FCOJ price variation but that still 80% remains unexplained.

---

[1] FCOJ futures contracts are for 15,000 pounds and at any given time nine to eleven contacts are outstanding with expiration schedules every second month (January, March, May, etc.), with at least two January months listed at all times.

The United States Department of Agriculture (USDA) orange production forecast issued in October has a particularly high impact on price volatility. The announcement date (the second week of October) accounts for more than 7% of the annual return variance (Boudoukh et al., 2007). The error of the forecast increases significantly in freeze years (mean error of 12.7%), but substantial variations in the forecast still occur regardless of freezes, which Boudoukh et al., 2007 attribute to "production uncertainty". The relationship between the FCOJ price volatility and the USDA forecast prompted us to question how a prediction of the direction and magnitude of the error of the forecast could affect the decision making process of multiple FCOJ market participants (growers, juice processors, and commodity investors, hedgers, traders and speculators), and if the "production uncertainty" of the forecast could be reduced by incorporating other climate variables. Therefore, unlike other analyses, we do not attempt to explain the FCOJ price volatility over time, but rather we focus exclusively on the effect of the USDA citrus forecast by developing models to predict its error using climate variables. To achieve this, we model the effects of past extreme events and diseases, and we assess the impact of different climate variables using local polynomial (locfit) models (Loader, 2013). The models developed open up the opportunity to assess the application of the resulting probabilistic forecasts of the USDA production forecast error on the trading decisions of the different FCOJ stakeholders, and to consider the inclusion of climate predictors in the USDA forecast.

**1.1 Orange production, the USDA forecast and its error**

The orange production in Florida has been declining over the last two decades (Figure 1), potentially attributable to significant extreme weather shocks and disease outbreaks. The strong hurricanes during 2004 and 2005 (namely, Charlie, Frances, and Wilma) greatly affected the orange production, almost to the same degree as the devastating freezes during the 1980's. Disease outbreaks have also played an important role in the production decline. The citrus greening disease (known as Citrus Huanglongbing), which causes premature fruit drop and higher tree mortality, has affected orange production in Florida since 2005 when it was first spotted (Spreen and Baldwin, 2014). It was estimated that without the citrus greening, the cumulative total of orange juice production from 2006 to 2011 would have been 951 million boxes compared to the actual production of 734.3 million boxes (Hodges and Spreen, 2012). The combination of extreme weather events, diseases and other fluctuations in climate has raised the error of the USDA October forecast. The production was significantly underestimated during 2004 and 2005, when the strong hurricanes hit, and again during 2013 and 2014 due to the effects of citrus greening. The error of the total orange production forecast ranges from a negative 1.3% to a positive 19.4% (Bouffard, 2015; USDA, 2015).

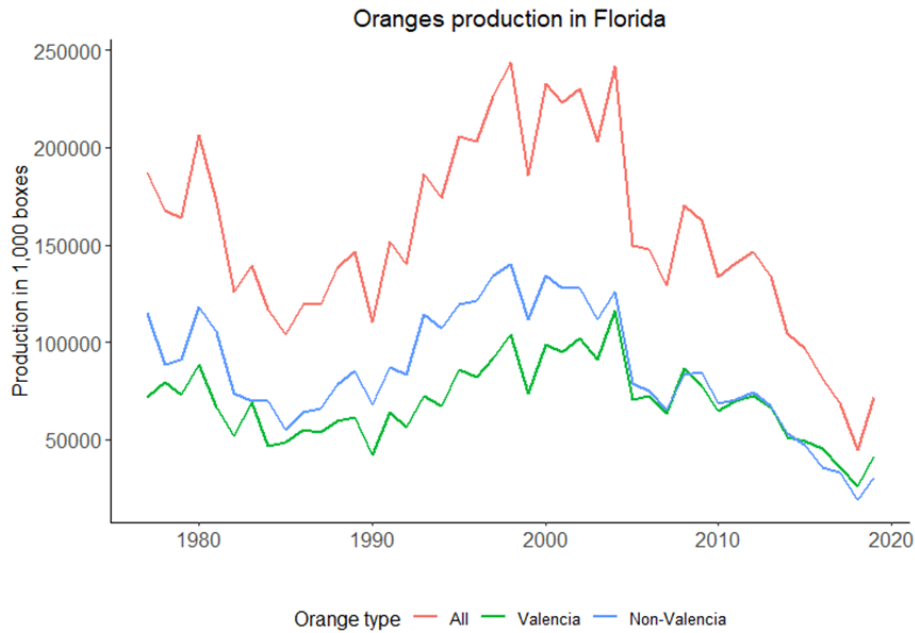

**Figure 1** Florida's orange production (per 1000 boxes). Data from the USDA orange production forecasts

The USDA citrus production forecasts are announced the second week of each month from October to July, and provide a statewide estimate of different types of citrus fruit, including two classes of oranges that are grouped as non-Valencia, and Valencia. The forecasts are based on physical monthly surveys, and the first announcement of the season (October) is conditioned on a non-freeze season; if a freeze occurs later in the season subsequent forecasts take into account the damage and its effects on production (Boudoukh et al., 2007). The forecast predictors are: the number of bearing trees, the number of fruits per tree, the percent remaining fruit at harvest (calculated based on estimated drop rates from monthly surveys) and the number of fruits per box. Non-Valencia oranges are harvested between November/December and February/March, and Valencia oranges are harvested between March/April and May/June. These forecasts do not include climate variables that can affect orange yield, but our models of the USDA forecast error use them as predictors. Due to the differences in the maturation and harvesting time, weather affects Valencia and non-Valencia oranges differently. Therefore, separate models were developed to predict the forecast error of each orange variant.

Climate variables occurring before the October forecast is issued (referred to as pre-forecast predictors hereafter) were evaluated to see if there is an effect on the forecast that the physical surveys are not capturing, and post-forecast climate predictors were used to assess the impact of future weather once the forecast has been issued. The pre-forecast predictors were used to build a statistical model to estimate the USDA forecast error at the time the October forecast is issued. The analysis of the relationships between post-forecast predictors and %Error were used to get a qualitative estimation of the direction of the forecast error using seasonal climate outlooks provided by the National Oceanic and Atmospheric Administration (NOAA). An example of the potential application of the pre-forecast and post-forecast error predictions in FCOJ trading is presented in the results section.

## 2. Data and Methods

Historical October forecast and production data for Valencia and non- Valencia oranges was obtained from individual USDA/NASS reports from 1977 to 2019. The non-Valencia forecast didn't include production of temples before 2006 so the Non-Valencia forecast from 1977 to 2005 was adjusted by adding the temple's forecast in those years to obtain a coherent time series.

Daily precipitation and temperature data from 1977 to 2019 were obtained from the National Climatic Data Center (NCDC). The stations covered counties with the highest orange production in Florida: Polk, Highlands, Hardee, De Soto, Manatee, Hillsborough, Charlotte, Glades, Hendry, Lee, Collier, Indian River, Martin, St. Lucie and Lake. One station was selected per county based on the completeness of its record.

The error of the forecast was calculated as the difference between the prediction made by USDA in October and the final production registered in the following July (Eq. 1); therefore a positive error indicates an over estimation of the production and a negative error an underestimate.

$$\% Error = \frac{F_{Oct(y-1)} - P_{Jul(y)}}{P_{Jul(y)}} * 100 \qquad [1]$$

where $y$ is the year. $F_{Oct(y-1)}$ is the forecast issued in October, and $P_{Jul(y)}$ the final production in July of the following year.

The resulting time series of Valencia and non-Valencia %Error are shown in Figure 2. The highest %Error (>10%) is related to the early 80's freezes, the 2004-2005 hurricanes, and the presence of citrus greening disease. The %Error is different for Valencia and non-Valencia oranges, and it is generally smaller for non-Valencias than it is for Valencias. This could be expected because the October forecast has a shorter lead-time for non-Valencias than for Valencias.

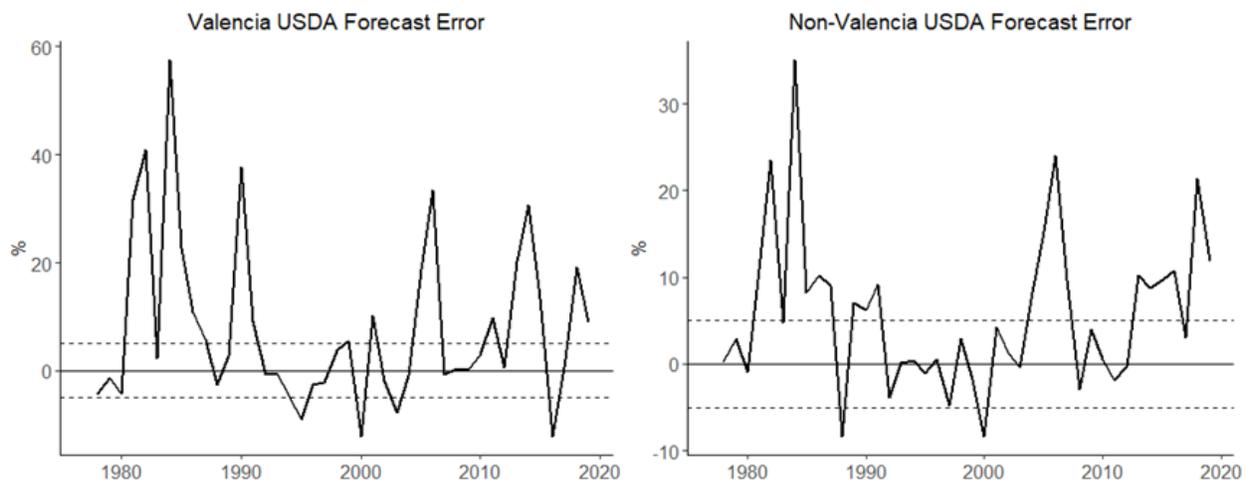

**Figure 2** Percent difference of forecast versus production for Valencia (left) and Non-Valencia oranges (right). The dotted lines indicate +/- 5% error.

To account for the differences in the forecast error caused by known extreme events and diseases, on the first step, a linear regression, which included a delta function placed on the known years for the extreme events and diseases was used. This way we could separate the effects of extreme events from other climate variables that affect the crop. The model had the following form of Eq.2:

$$\% Error = \Delta_{Freezes} + \Delta_{Hurricanes} + \Delta_{Cg} + \varepsilon \quad [2]$$

$$\% Error(y) = \alpha_1 \Delta_{Freezes}(y) + \alpha_2 \Delta_{Hurricanes}(y) + \alpha_3 \Delta_{Cg}(y) + \varepsilon(y)$$

where, $\Delta_{Freezes}(y), \Delta_{Hurricanes}(y), \Delta_{Cg}(y)$ is equal to 1 if the associated event occur on the year $y$, and is equal to 0 otherwise.

The variables treated this way included freezes, hurricanes, and the presence of citrus greening (Cg) starting in 2013 to date (the drop rates assumed by the USDA to issue the forecasts have not been correct since 2013 when the disease started to have a broader impact in the State). We use the residual $\varepsilon$, which subdivided the effect of the three extreme events, to investigate their impact from other climate predictors.

## 2.1 Climate predictors

Weather prior to and after the USDA forecast issued was considered to obtain climate predictors for the models because perennial trees such as oranges are affected by weather throughout the year (Valiente and Albrigo, 2004; Oswald; Strategic Planning for the Florida Citrus Industry, 2010; Morton, 1987). Different relationships with weather are observed during dormancy (before blooming), bloom, growth, and maturation and in some instances they are opposite; for example, cool temperatures in the winter during dormancy are related to high yield while the same condition during growth and final maturation can have detrimental effects.

For the screening of potential climate predictors, the following variables were calculated: the number of days in a month when the minimum temperature is below 1°C (month1C), below 4°C (month4C), and the number of consecutive days when the mean temperature is below 7°C (monthc) to reflect cold spells because prolonged cold temperatures during maturation can reduce orange yield (Morton, 1987). These variables were obtained for November, December, January, February, and March. Variables related to precipitation were also included even though they are believed to have a milder effect in orange yield than temperature in Florida because practically all fields are irrigated. However, an excess rainfall during the spring bloom is linked to the appearance of diseases and to the yield losses (USDA, 1998). The number of days exceeding the 75[th] daily rainfall percentile from February to April (FMAQ75) was added to the potential predictor mix (calculated taking all data, and confirming that the 75[th] percentile was not equal to 0). Other variables related to precipitation were the total rainfall from June to August (JJA), and the summer water deficit based on theoretical water needs for oranges (Kisekka et al.; Devineni et al, 2013).

Two types of predictor approaches were evaluated. The first used predictors calculated with data from the individual stations at the county level. In the second one, weather stations were grouped into 4 clusters and the mean across the stations within a cluster was taken to be the predictor. The clustering approach was done on the historical time series of orange yield of the counties based on Kmeans (MacQueen, 1967). The resulting "climate clusters" were identified according to the yield characteristics of each region, assuming that weather in each cluster would affect yield in the clusters in a similar fashion, and that grouping could reduce the noise of data from individual stations. The clustering pattern obtained was very similar to the official commercial citrus production areas of the USDA/NASS; Cluster 1(C1): Manatee, Hardee and De Soto (western region); Cluster 2 (C2): Polk, and Highlands (central region); Cluster 3 (C3): Charlotte, Glades, Collier, and Hendry (southern region); Cluster 4 (C4): St. Lucie, and Indian River (Indian River District).

Predictors from individual stations and clusters were selected using common predictor selection techniques such as correlation maps (Pearson, and Spearman), mutual information, and scatter plot analyses between the climate variables at each county and the %Error residuals for Valencias and non-Valencias, after the hurricane, freezes and disease effects were removed ($\varepsilon$ in Eq.2). Another strategy for predictor selection when the pool of potential candidates is large could have been to use the least absolute selection and shrinkage operator (Lasso) model (Hastie et al. 2001), as described by Lobell et al., 2011, but it is limited to linear responses. The number of predictors retained in the model was constrained by having 39 years of forecast data available to avoid over fitting caused by using too many predictors.

The following modeling approaches were evaluated: generalized linear models, k-NN (Lall and Sharma, 1996), and multivariate local regression using the R package locfit (Loader, 2013). The best cross validated results were obtained with locfit. The underlying model is

$$Y_i = \mu(x_i) + \varepsilon \qquad [3]$$

where, $Y_i$ is the residual $\varepsilon$ of Eq (2)

The function μ(x) is estimated by fitting a polynomial model (usually linear or quadratic) within a sliding window so for each fitting point x, a locally weighted least square criterion is considered (Loader, 2013). The model selection was done using the generalized cross validation statistic (GCV) included in the locfit package, and was evaluated for each model using different smoothing parameters alpha (Loader, 2013), which determines the nearest-neighbor based bandwidth (i.e. an alpha=0.5 covers 50% of the data).

Three locfit models were evaluated for each type of orange; the first two used pre-forecast station-predictors and cluster-predictors, and the third used post-forecast cluster-predictors. The latter serves only for qualitative assessments of the likelihood of a certain error based on the seasonal climate outlooks from NOAA. The models were calibrated with data from 1977 to 2014.

## 2.2 FCOJ and the USDA October Forecast

The next step was to evaluate the relationship between the announcement of the forecast and the FCOJ returns. Milonas et al., (1987) and Boudoukh et al., (2007) found that the USDA October forecasts are relevant for farmers, processors, traders, and speculators despite the availability of satellite data and privately generated forecasts, therefore, we expected to see an increase in volatility in the days close to the announcement. We confirmed this by analyzing the volatility of the daily returns of the closest to maturity contract throughout the year, and on the day before, the day of and the day after the forecast is issued (refer to the supplemental material). In all cases, there were unusual spikes on the returns in the second week of October (when the forecast is issued), particularly on the day before and the day of the announcement. We also evaluated the possibility of predicting the magnitude and direction of the returns using historical information of the forecasts, the latest orange production, and the closing prices adjusted to inflation as predictors, but we didn't find any significant predictability. This would have provided a better quantitative tool for trading once the percent error of the forecast was predicted, but we could still propose a qualitative decision making framework of the potential actions of the stakeholders once the prediction of the error is issued.

## 3. Results

The %Error residuals for non-Valencia had strong positive correlations with JJA precipitation and with Jan4c in station and cluster pre-forecast predictors. In the station-predictor approach JJA precipitation in Hardee is significant, and this was also observed in the cluster-predictor approach where C1 JJA precipitation (which includes Hardee) was the strongest predictor. Likewise, the Jan4c predictors are consistent in both approaches; Collier at the station level and C3 (which includes Collier) at the cluster level. The % Error for Valencia oranges however did not present a consistent behavior between the pre-forecast station-predictors and the cluster-predictors. In the station pre-forecast approach, Valencia's residuals were negatively correlated with extreme rainfall days in May (MayQ75) in Indian River only. When the climate predictors were clustered, the %Error residuals for Valencia had significant positive correlations with the maximum length of consecutive days under 7ºC (Febc) in C2, and with extreme rainfall from February to April (FMAQ75) in C1; no significant correlations with MayQ75 were found in any cluster. For the post-forecast cluster predictors, Dec4c had the strongest correlations with Valencia and non-Valencia in C3 and C2 respectively.

Once the potential predictors were identified, the possible combinations of locfit models (including citrus greening, freezes, and hurricanes) to predict the %Error were evaluated using an ANOVA table computing GCV (Table 1 to Table 6).

The probability density functions of the forecast can be obtained from the models selected using a bootstrap resampling approach, such as the one followed by Lall and Sharma (1996), but adapting it to locfit. The resulting probabilistic forecast can be used to evaluate decisions under the different likely scenarios (more about this on the Discussion section.

### 3.1 Non- Valencia oranges locfit models

From Table 1, the variable with the most impact on the error of non-Valencia using the pre-forecast cluster-predictor approach is freezes, followed by hurricanes. From that analysis, JJA precipitation does not contribute to a reduction in GCV but Jan4c does. Therefore, the model selected was:

$$locfit(\% Error = \Delta_{Freezes} + \Delta_{Hurricanes} + C3_{Jan4C}) \quad [4]$$

With alpha=0.65.

The same conclusion was reached from the individual station approach (Table 2), where the best model only includes Jan4c in Collier (which is in C3) and not JJA precipitation in Hardee:

$$locfit(\% Error = \Delta_{Freezes} + \Delta_{Hurricanes} + Collier_{Jan4C}) \quad [5]$$

With alpha=0.65.

GCV for the best model is slightly lower in the station-predictor model than in the cluster-predictor model (34.2 versus 41.3).

Attempting to explain potential causal relationships, C3 includes some of the biggest non-Valencia orange producers so extreme weather in this area is likely to have an impact in the overall orange production and therefore, in the forecast error. The positive correlation between Jan4c and the %Error of the forecast means that if Jan4c is large (very cold temperatures in January), then the forecast is likely to over predict. Flower intensity is enhanced with greater accumulation of hours at low temperatures between 11-15ºC in Florida (Valiente and Albrigo, 2004) but García-Luis et al. (1992), found that exposure of the shoots to 4ºC does not induce flowering and that it actually reduces flowering of partially induced buds; this observation could partially explain the tendency of the forecast to over predict when there are many days below 4ºC in January.

The most relevant post-forecast climate predictor for the error found was Dec4c in the central region C2 which has the biggest orange producers (Table 3); this is in agreement with the observation that exposure to freezes/very cold temperatures in the final maturation months can harm orange crops.

**Table 1** Non-Valencia cluster GCV considering different predictor combinations (pre-forecast predictors).

| Freezes | Hurricanes | JJAprecip Clus1 | Jan4c C3 | GCV |
|---------|------------|-----------------|----------|-------|
| 1 | 1 | 1 | 0 | 49.9 |
| 1 | 1 | 0 | 1 | **41.3** |
| 1 | 0 | 1 | 1 | 81.5 |
| 0 | 1 | 1 | 1 | 108.5 |
| 1 | 1 | 1 | 1 | 61.4 |

**Table 2** Non-Valencia individual GCV considering different predictor combinations (pre-forecast predictors).

| Freezes | Hurricanes | JJAPrecip Hardee | Jan4c Collier | GCV |
|---|---|---|---|---|
| 1 | 1 | 1 | 0 | 39.4 |
| 1 | 1 | 0 | 1 | **34.2** |
| 1 | 0 | 1 | 1 | 73.2 |
| 0 | 1 | 1 | 1 | 75.6 |
| 1 | 1 | 1 | 1 | 42.1 |

**Table 3** Non-Valencia cluster GCV considering different predictor combinations (post-forecast predictors).

| Freezes | Hurricanes | Dec4cC2 | GCV |
|---|---|---|---|
| 1 | 0 | 1 | 66.5 |
| 0 | 1 | 1 | 84.9 |
| 1 | 1 | 1 | **51.4** |

### 3.2 Valencia oranges locfit models

The ANOVA table for Valencia's % Error using the pre-forecast cluster-predictor approach (Table 4) shows that freezes explain the greatest variance, followed by citrus greening, and hurricanes; the addition of Febc from C2 increases the GCV score so the best model only includes FMAQ75 from C1:

$$_{locfit}(\%\ Error = \Delta_{Freezes} + \Delta_{Hurricanes} + \Delta_{Cg} + C1_{FMAQ75}) \tag{6}$$

With alpha=0.70.

In the individual stations pre-forecast approach (Table 5) the order of significance of freezes, citrus greening and hurricanes remains the same as in the cluster-predictor models, and the inclusion of MayQ75 in Indian River (IR) reduces the GCV. Therefore, the individual model for Valencia %Error is:

$$_{locfit}(\%\ Error = \Delta_{Freezes} + \Delta_{Hurricanes} + \Delta_{Cg} + IR_{MayQ75}) \tag{7}$$

With alpha=0.70.

The GCV score for the individual stations model is lower than in the cluster model (53.4 versus 73.8).

The cluster-predictor FMAQ75 could potentially be justified because the amount of water in those months influences the type of inflorescence and water stress can lead to heavy drop of fruit in June (Carr, 2013). The inflorescence can be leafy or leafless and the degree of fruit set is different for each type (Carr, 2013). A leafless inflorescence is less likely to set fruit; there may be many flowers but just a small proportion becomes fruit (normally only around 1% of flowers reaches maturity; Iglesias et al, 2007). Therefore, having plentiful rain in the months from February to April (FMAQ75) should lead to an enhancement of flowering and potentially to more fruit production (having small June drop values), but the control of fruit set later in the season involves many factors which relationships are not yet fully understood (Carr, 2013). This can lead to an exceedingly positive expectation of production, and therefore to an over-estimation.

In the station-predictor model, the relationship with MayQ75 in Indian River is surprising since the county is a small producer of Valencias and would not be expected to have a high weight in the production forecast error. The causal explanation of the predictor could be similar to behavior of FMAQ75 in the cluster model, but the weight in the model is hard to justify with causal arguments.

**Table 4** Valencia cluster GCV considering different predictor combinations (pre-forecast predictors).

| Cg | Freezes | Hurricanes | FebcClus2 | FMAQ75C1 | GCV |
|----|---------|------------|-----------|----------|------|
| 1  | 1       | 1          | 0         | 1        | **73.8** |
| 1  | 1       | 1          | 1         | 0        | 77.7 |
| 0  | 1       | 1          | 1         | 1        | 193.1 |
| 1  | 1       | 0          | 1         | 1        | 171.8 |
| 1  | 0       | 1          | 1         | 1        | 232 |
| 1  | 1       | 1          | 1         | 1        | 75.6 |

**Table 5** Valencia individual GCV considering different predictor combinations (pre-forecast predictors).

| Cg | Freezes | Hurricanes | MayQ75 Indian River | GCV |
|----|---------|------------|---------------------|------|
| 1  | 1       | 0          | 1                   | 106.4 |
| 0  | 1       | 1          | 1                   | 105.2 |
| 1  | 0       | 1          | 1                   | 229.9 |
| 1  | 1       | 1          | 1                   | **53.4** |

For Valencia oranges, the most important post-forecast climate predictor of the error was Dec4c in C3 (Table 6), the second largest producing cluster of Valencias. Just like for non-Valencias, an over estimation of the production of Valencias is more likely when oranges are exposed to temperatures below 4ºC in the final maturation stage.

**Table 6** Valencia cluster GCV considering different predictor combinations (post-forecast predictors)

| Cg | Freezes | Hurricanes | Dec4cC3 | GCV |
|----|---------|------------|---------|------|
| 0  | 1       | 1          | 1       | 127.6 |
| 1  | 1       | 0          | 1       | 128.2 |
| 1  | 0       | 1          | 1       | 269.5 |
| 1  | 1       | 1          | 1       | 75.5 |

**4. Discussion: The conceptual use of the USDA forecast error prediction model in FCOJ trading**

Farmers can hedge or secure a price to protect their crops either: i) by buying a put option contract, which gives the farmer the right to sell at a specified "strike" price at maturity when the crop is ready, or, ii) by taking a short position on FCOJ futures contract at the time of the option expiry if it is "in-the-money" (the strike price is higher than the market price). To buy the put option, they have to pay an option premium. The FCOJ buyer side can secure a price by buying a call option at a specified strike price and expiry date. If at the time of expiry, the contract is in-the-money, the FCOJ processor has the right to take a long orange futures position. So for a FCOJ put option, if the market spot price is higher, the contract is worthless (the farmer loses the premium but can still sell at the market spot price), and for a call option when the market price is lower, the contract is worthless (the orange juice processor loses the premium but can still buy at market price). For a speculator, the price discovery trade is more straightforward; if the trader believes the price of FCOJ will rise, the choice is to "go long" a futures contract (buy futures contract); otherwise if the belief is that the price will fall, the choice is to "go short" (sell contract) a futures contract (for more information refer the ICE FCOJ trading brochure). Following this logic in the context of the effect the USDA production forecast should have on the different types of buyers and sellers, the decision making process should follow similar guidelines as the ones presented in Table 7 for farmers and juice processors (hedgers) once the direction and magnitude of the forecast error is known.

Under the hypothetical situation where the cumulative density function of the prediction of the USDA forecast error for a given year is the one shown for non-Valencia oranges in Figure 3 -left, the decision maker could take different approaches according to their degree of risk aversion or acceptance, and their position. In this example, the probability of the error being less than 5% (regardless if the error can go in the negative direction) is around 0.07, which means that there is 7% chance that the USDA forecast error would be between -1 to 5% (the probability of being above 5% is therefore 0.93 or 93%). In this scenario an orange juice processor may consider that a 93% chance of the USDA forecast overestimating production by 5% is too high and they need to secure supply (specially for processors that do not own citrus land), then they would opt for a call option (taking a long position) as in Scenario A to protect themselves from price and supply shocks. In the case of a citrus grower, under the same scenario she can decide that the chances of the USDA over predicting the production by a great extend are very high, and therefore believe that the price of oranges can increase. The growers may then also opt to take a long position. On the other hand, in the scenario that the forecast was that shown in Figure 3 for Valencia oranges (right), the growers would opt to protect themselves by getting a put option (take a short position) as in Scenario B, because the chances of an over-prediction are very low and the price could go down. In all cases, the decision tree of the different stakeholders will vary according to their risk perception based on the probability distribution of the prediction of the forecast error.

The direction of the error of the forecast can also be weighed against post-forecast climate outlooks as presented in the Data and Methods section. In the Results section it was concluded that the post-forecast climate condition that is most likely to cause an over estimation for both orange types is a high number of days below 4ºC (Dec4c). The NOAA seasonal outlooks are divided in four categories: above normal, below normal, normal and equal chances of falling in any of those categories. So for example, if NOAA's outlook predicts that temperature in Florida in December will be below normal, the chances of an overestimation of the orange production by the USDA increase. In that case the decision process would be as Scenario A in Table 7. This additional piece of information can then modify the decision taken solely based on the probabilistic estimate of the error based on pre-forecast climate.

Table 7 Trading scenarios in relation to the USDA forecast.

| Scenario | Impact in Price | Farmer | Processor |
|---|---|---|---|
| A – The forecast will over-estimate production (positive error) P<x* | Market price in the future might be higher than the current strike price. | Take a long position in put option or wait to sell at the spot market. Sell put options. There could be a loss of premium for put option buyers but they can sell at market price | Need to secure supply by placing more futures contracts (taking long position) or buying call options before the price increases. |
| B – The forecast will under-estimate production (negative error) P>x | Market price in the future might be lower than the current strike price. | Buy put options to secure the crop price. Take short position in futures contract | Buy enough contracts to secure a minimum production threshold but leave some of the supply to the spot market. Sell call options (loss of premium for call option buyer). Take short position in futures contracts. |
| C – The forecast is close to the production P~x | Market price in the future might be similar to the strike price. | Buy put options to secure the crop price. Take short position in futures contract or wait to sell at market price. | Safe scenario for procurement, buy "safety supply" as contracts and the rest in the spot market |

*P= production, x=forecast

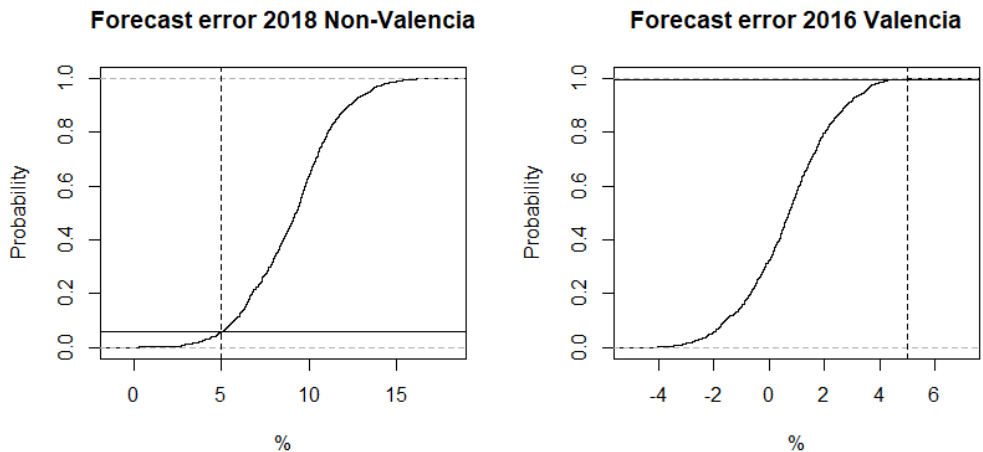

**Figure 3** Cumulative density function of a hypothetical prediction of the USDA forecast error for a given year.

To demonstrate the cases of the losses and gains that processors and farmers could experience, we obtained the closing price from 1977 to 2014 of FCOJ on the day the forecast is issued and computed a monthly average of the prices of nearest to maturity contracts in March and May, the end of non-Valencia and Valencia harvesting seasons respectively (spot/cash prices tend to converge with futures prices during the delivery month, so we considered appropriate to use those prices). The gain/loss of the contract (depending on the position taken) was calculated by subtracting the price of the FCOJ future the day of the forecast from the nearest to maturity contract in March and May. The positive values in this case correspond to price increases; therefore, to gains for long positions in futures contracts. The negative values on the other hand correspond to gains for short positions. The time series of the positive and negative values were bootstrapped 1000 times computing the medians in each iteration; the absolute mean value of the sampled medians was taken as the expected gains for the short and long positions ($E_{short}$ and $E_{long}$ respectively). $E_{long}$ for non-Valencia oranges was 20.4 cents/pound ($3,060.4 per 15,000-pound contract) and for Valencia was 26.32 cents/pound ($3,947.5 per contract). $E_{short}$ was 13.08 cents/pound ($1,963 per contract) for non-Valencia and 18.74 cents/pound ($2,811.1 per contract) for Valencia. Then, taking the probability distribution of the error of the USDA forecast in a given year, such as those in Figure 3, we can set decision rules. In this example we are choosing 5% of the forecast error as a parameter indicating a potential increase in price (overestimation of supply) and anything below 5% as an indicator that prices will remain similar to the current price or that they will decrease (underestimation of supply). With this rule, we follow the decision tree shown in Figure 4 for Valencia oranges to calculate the expected monetary values (EMV) of each branch. In the 2018 season, the USDA forecast issued in October 2017 had an error for Non-Valencia oranges of 21.4%. In that year our model predicted that production was likely to be lower than the USDA forecast, with a mean estimate of 9.1%, and a probability of exceeding 5% of 0.93 (Figure 3). In this particular case, the most favorable decision is to take a long position (or wait to sell in the spot market). In the other hand, in the season of 2016-2017 the error of the USDA forecast for Valencia oranges was -12.1% so the production was under-predicted. That year our forecast predicted that the production of Valencia oranges was likely to be higher than the USDA forecast, with a probability of .003 of exceeding a 5% error, in that case, the decision tree would indicate that taking a short position is better.

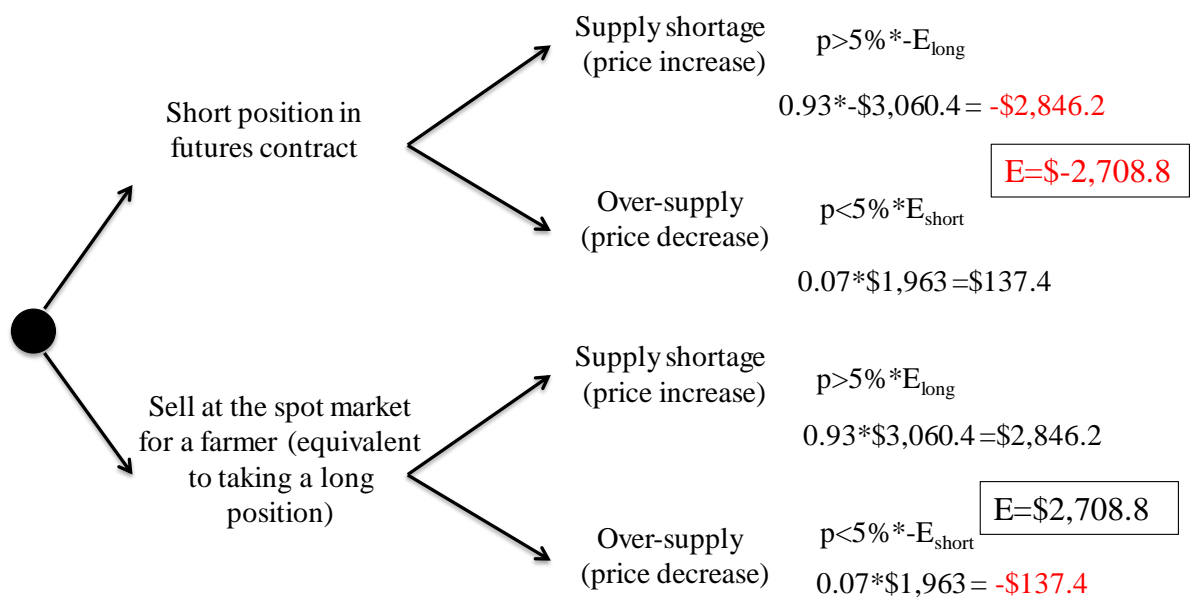

**Figure 4** Decision tree using the 2018 forecast error probabilities for non-Valencia oranges.

However, in reality the process is more complicated. For example, for orange juice processors, Roll (1984) found that some of the major companies in the orange juice business did not have a relation with FCOJ returns. This was explained because some of them own citrus land and therefore they could offset or hedge their own supply by selling OJ futures. This in theory makes them more susceptible to demand shocks than to supply shocks, depending on the volume of citrus produced in their own fields and the extent of the shock. Therefore, the decision framework that we present would have to be tailored for each participant of the orange market according to his or her specific situation and risk profile, but the prediction of the percent error could be used by all of them.

A methodology to develop models to predict the error of the October USDA orange production forecast was presented. The models incorporate climate parameters from before and after the forecast is issued and provide a tool for decision makers participating in the orange and FCOJ markets. Access to orange production data, and forecast outputs are available as an app in https://paupak4.shinyapps.io/ovt2/

## Acknowledgments

We would like to thank PepsiCo for their funding. We also thank the USDA-NASS and Dr. Gene Albrigo for providing historical orange production and forecast information.